\documentclass[a4paper]{jpconf}
\usepackage{graphicx}

\newcommand{\apss}{{Ap\&SS}}
\newcommand{\bain}{{Bull. Astron. Inst. Netherlands}}
\newcommand{\actaa}{{Acta Astron.}}
\newcommand{\zap}{{ZAp}}
\newcommand{\aj}{{AJ}}
\newcommand{\aap}{{A\&A}}
\newcommand{\apj}{{ApJ}}
\newcommand{\mnras}{{MNRAS}}

\begin{document}
\title{The Photometric Investigation of V921 Her using the Lunar-based Ultraviolet Telescope of Chang'e-3 mission}

\author{Xiao Zhou$^{1,2,3}$, Sheng-Bang Qian$^{1,2,3}$, Jia Zhang$^{1,2}$, Lin-Jia Li$^{1,2}$ and Qi-Shan Wang$^{1,2,3}$}

\address{$^1$ Yunnan Observatories, Chinese Academy of Sciences (CAS), P. O. Box 110, 650216 Kunming, China\\
$^2$ Key Laboratory of the Structure and Evolution of Celestial Objects, Chinese Academy of Sciences, P. O. Box 110, 650216 Kunming, China\\
$^3$ University of Chinese Academy of Sciences, Yuquan Road 19\#, Sijingshang Block, 100049 Beijing, China}

\ead{zhouxiaophy@ynao.ac.cn}

\begin{abstract}
The light curve of V921 Her in ultraviolet band observed by the Lunar-based Ultraviolet Telescope (LUT) is analyzed by the Wilson-Devinney code. Our solutions conclude that V921 Her is an early type marginal contact binary system with an additional close-in component. The binary system is under poor thermal contact with a temperature difference of nearly $700K$ between the two components. The close-in component contributes about $19\,\%$ of the total luminosity in the triple system. Combining the radial velocity study together with our photometric solutions, the mass of the primary star and secondary one are calculated to be $M_1 = 1.784(\pm0.055)M_\odot$, $M_2 = 0.403(\pm0.012)M_\odot$. The evolutionary scenario of V921 Her is discussed. All times of light minimum of V921 Her available in the bibliography are taken into account and the $O - C$ curve is analyzed for the first time. The most probable fitting results are discussed in the paper, which also confirm the existence of a third component ($P_3=10.2$ year) around the binary system. The period of V921 Her is also undergoing a continuously rapid increase at a rate of $dP/dt=+2.79\times{10^{-7}}day\cdot year^{-1}$, which may due to mass transfer from the less massive component to the more massive one.
\end{abstract}

\section{Introduction}

In December 2013, the Chinese Lunar Exploration Program carried out the Chang¡¯e-3 mission \cite{2014RAA....14.1511I}, which sent a LUT to the surface of the moon.
LUT is the first robotic astronomical telescope deployed on the moon surface. It has been working for nearly two years since landing on the moon. Eclipsing binaries are among its observational targets. We have obtained some light curves (LCs) to be analyzed, and V921 Her is a such one.

The light variability of V921 Her (HIP 82344,$V$ = $9^{m}.49$) was first discovered by the Hipparcos mission \cite{1997ESASP1200.....E} and the Hipparcos Catalogue gave its spectral type to be A5.
Later, Ruci{\'n}ski et al. \cite{2003AJ....125.3258R} carried out the radial velocity study of V921 Her. The results determined that V921 Her was an A-subtype contact binary system with its spectral type to be A7IV. They obtained its mass ratio $(M_2/M_1)$ to be $q = 0.226(5)$. Compared with other W UMa type close binary systems of the same spectral type, V921 Her has a quite long period, which is P = 0.877366d. The first photometric analysis of V921 Her was done by Gazeas et al. \cite{2006AcA....56..127G}. They redetermined the mass ratio of the binary system to be $q = 0.244$ and set the effective temperature of the primary star to be $T_1=7700K$. Their results concluded that V921 Her was a shallow contact binary system. However, Karami \& Mohebi \cite{2007JApA...28..217K} and Karami et al. \cite{2009AN....330..836K} also obtained the mass ratio of V921 Her, both of which were consistent with those determined by Ruci{\'n}ski et al. \cite{2003AJ....125.3258R}.

Orbital period research is a very important part for close binary systems since the period variations contain information about the dynamic interaction occurring between the components of binary systems. However, the period change investigation of V921 Her has been neglected since it was discovered. In the present work, light curve of V921 Her observed by the LUT is analyzed with the Wilson-Devinney (W-D) code, and its period variations are investigated for the first time. The results will give us comprehensive understanding to the physical properties and dynamical evolution of this contact binary system.

\section{New Photometric Observations}
The LUT is a 150 mm, F/3.75 Ritchey-Chretien telescope working at a Nasmyth focus. It is equipped with a UV-enhanced back-illuminated AIMO 1k CCD with a field of view to be $1.3^\circ \times 1.3^\circ$, thus the CCD pixel scale is 4.7'' $pixel^{-1}$. The passband of filter is about 245-345 nm, with its peak at 250 nm \cite{2015RAA....15.1068W}.

The light curve of V921 Her was observed continuously in ultraviolet band by the LUT on February 02, 2015. PHOT subroutine (measured magnitudes for a list of stars) of the IRAF \footnote {The Image Reduction and Analysis Facility is hosted by the National Optical Astronomy Observatories in Tucson, Arizona at URL iraf.noao.edu.} aperture photometry package was used to reduce the observed images \cite{2015Ap&SS.358...47M}.

Using the least-squares method \cite{1956BAN....12..327K}, we obtained two times of light minimum through the light curve observed, which are listed in Table 1.
\begin{table}[!h]
\normalsize
\begin{center}
\caption{New CCD times of light minimum for V921 Her}\label{Newminimum}
\begin{tabular}{cccccc}\hline
    JD (Hel.)     &  Error (days)  & Min. \\\hline
  2457056.3349    & $\pm0.0015$    &   I  \\
  2457056.7733    & $\pm0.0019$    &   II \\
\hline
\end{tabular}
\end{center}
\end{table}

The phase of our observations are calculated with the following linear ephemeris:
\begin{equation}
Min.I(HJD)=2457056.3349(15)+0^{d}.877379\times{E}\label{linear ephemeris}
\end{equation}

\section{Investigation of the Orbital Period}
The orbital period variations of a binary star reveal dynamic interactions between its components or the gravitational interactions between the binary system and the close-in component orbiting around it. During this work, all times of light minimum of V921 Her available in the bibliography are taken into account and listed in Table 2. All of the minimum data are observed with CCD camera.  The $(O-C)$ (observed times of light minimum - calculated times of light minimum) values of all light minima are calculated with the linear ephemeris below,
\begin{equation}
Min.I(HJD) = 2448500.1250+0^{d}.877379\times{E}\label{linear ephemeris}
\end{equation}
The Epoch and corresponding $O-C$ values are listed in the third and fourth columns of Table 2, respectively. The origin in time used in Equation (2) is the first minimum light available in the bibliography for V921 Her.

As shown in the upper panel of Fig. 1, the linear ephemeris in equation (2) can not explain the $O - C$ variations efficiently. We consider long term increase (or decrease) and periodic variations superposed on the linear ephemeris, then
\begin{equation}
O - C = \Delta T_0 + \Delta P_0\times E +\frac{1}{2} \frac{dP}{dE}\times E^2 + \tau
\end{equation}
where $\Delta T_0$ and $\Delta P_0$ are corrections to the initial ephemeris and period,  $\tau$ is the periodic element given by Irwin (1952) \cite{1952ApJ...116..211I} in the case of $e = 0$.
Based on the least-squares method, the most probable results of $(O-C)$ curve fitting are given out.

\begin{table}[!h]
\caption{$(O-C)$ values of light minimum for V921 Her.}\label{Minimum1}
\begin{center}
\normalsize
\begin{tabular}{cclrlcc}\hline\hline
JD (Hel.)      &  Min &   Epoch     & $(O-C)$      &   Error       &  Reference       \\
(2400000+)     &      &  (cycles)   &  (days)      &  (days)       &\\\hline
48500.1250     & I    &  0	        & 	      0    &               &  1    \\
51318.2680     & I    &  3212       &    0.0017    &               &  2    \\
52790.5033     & I    &  4890	    & 	-0.0050    &    0.0001     &  3    \\
52808.4900     & II   &  4910.5     &   -0.0046    &    0.0008     &  4    \\
52840.5150     & I    &  4947	    & 	-0.0039    &    0.0005     &  5    \\
52862.4454     & I    &  4972       &   -0.0080    &    0.0013     &  5    \\
53087.9381     & I    &  5229	    &   -0.0017    &    0.0002     &  6    \\
53095.3998     & II   &  5237.5     &    0.0023    &    0.0025     &  7    \\
53131.8020     & I    &  5279	    &   -0.0067    &    0.0004     &  8    \\
53146.7240     & I    &  5296	    &   -0.0002    &    0.0003     &  8    \\
53463.4610     & I    &  5657	    &    0.0030    &    0.0001     &  3    \\
53821.8693     & II   &  6065.5	    &    0.0020    &    0.0003     &  9    \\
55270.8510     & I    &  7717	    &   -0.0077    &    0.0002     &  10    \\
55298.9376     & I    &  7749	    &    0.0027    &    0.0003     &  11    \\
55374.3931     & I    &  7835     	&    0.0036    &    0.0005     &  12    \\
57056.3349     & I    &  9752	    &    0.0099    &    0.0015     &  13    \\
57056.7733     & II   &  9752.5     &    0.0096    &    0.0019     &  13    \\
\hline
\end{tabular}
\end{center}
\textbf
{\footnotesize Reference:} \footnotesize (1) Hipparcos mission \cite{1997ESASP1200.....E}; (2) ROTSE project \cite{1997AAS...191.4815M}; (3) Pribulla et al. \cite{2005IBVS.5668....1P}; (4) Br{\'a}t et al. \cite{2007OEJV...74....1B}; (5) Pejcha \cite{2005IBVS.5645....1P}; (6) Nelson \cite{2005IBVS.5602....1N}; (7) Krajci \cite{2005IBVS.5592....1K}; (8) Dvorak \cite{2005IBVS.5603....1D}; (9) Nelson \cite{2007IBVS.5760....1N}; (10) Dvorak \cite{2011IBVS.5974....1D}; (11) Nelson \cite{2011IBVS.5966....1N}; (12) Br{\'a}t et al. \cite{2011OEJV..137....1B}; (13) present work.
\end{table}

The resulting ephemeris is
\begin{equation}
\begin{array}{lll}
Min. I=2448500.1299(\pm0.0004)+0^{d}.8773763(\pm0.0000001)\times{E}
         \\+3.35(\pm0.09)\times{10^{-10}}\times{E^{2}}
         \\+0^{d}.0052(\pm0.0002)\sin[0^{\circ}.08475\times{E}+250^{\circ}(\pm1^{\circ})]
\end{array}
\end{equation}
With the quadratic term included in this ephemeris, a continuous period increase, at a rate of
$dP/dt=+2.79(\pm0.08)\times{10^{-7}}day\cdot year^{-1}$ is determined. The sinusoidal term reveals a cyclic period change with a period of 10.2 years and an amplitude of 0.00516 days. The residuals from equation (3) are showed in the lower panel of Fig. 1.

\begin{figure}[!h]
\centering
\includegraphics[width=1.0\columnwidth]{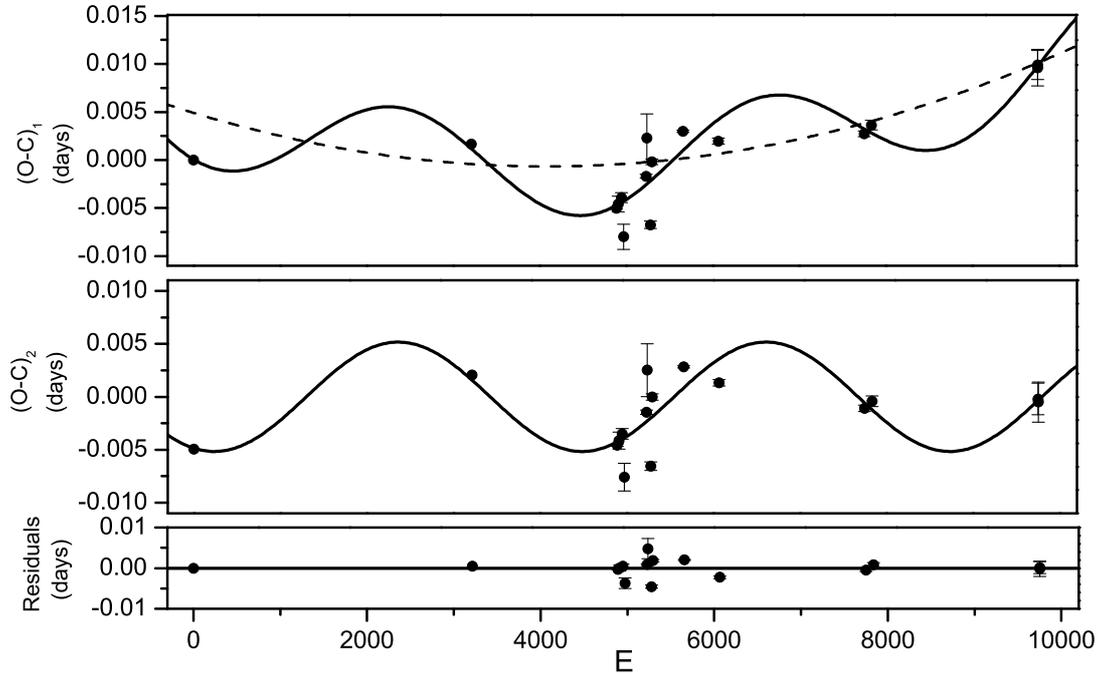}
\caption{\small The $(O-C)_1$ values of V921 Her from the linear ephemeris of Equation (2) are presented in the upper panel. The solid line in the panel refers to the combination of a upward parabolic variation and a cyclic change. The dashed line represents the upward parabolic variation which reveals a continuous increase in the orbital period. In the middle panel, the $(O-C)_2$ values with the quadratic part in Equation(3) removed are displayed, where a cyclic change is more clear to be seen. After both the parabolic change and the cyclic variation are removed, the residuals are plotted in the lowest panel.}
\end{figure}

It has to be mentioned that the data point of E = 7717 is not used since it is apparently inconsistent with the following two data point. The large group of minima are between E=4800 and E = 6000 cycles, the rest consists of only 6 very small groups of minima widely dispersed over time. To test the evolution of the O-C diagram, more and more data are necessary in the near future.

The orbital period of binary system is a very important parameter. In the present paper, comparing the period published by other researchers and databases, we chose the initial period to be $P = 0^{d}.877379$, since there is only a very small difference among them. The revised period in Equation (4) is 0.8773763(1), which confirms that the period we used is acceptable.

\section{Analysis of the Light Curves}

To understand the geometrical structure and physical properties of V921 Her, the observed light curve is analyzed using the W-D code (Version 2013) \cite{Wilson1971,Wilson1990,Wilson2010}.
According to the spectral type (A7IV) and mass ratio ($q=0.226$) derived by Ruci{\'n}ski et al. \cite{2003AJ....125.3258R}, the effective temperature of star 1 (star eclipsed at primary minimum light) is set to be $T_1=7700K$ \cite{Cox2000}, and the mass ratio is fixed to $q=0.226$. Radiative and convective outer envelopes are assumed for star 1 and star 2, respectively. Thus,  The bolometric albedo $A_1=1, A_2=0.5$ \cite{1969AcA....19..245R} and the values of the gravity-darkening coefficients $g_1=1, g_2=0.32$ \cite{1967ZA.....65...89L} are used. To account for the limb darkening in detail, logarithmic functions are used. The corresponding bolometric and passband-specific limb-darkening coefficients are chosen from Van Hamme's table \cite{1993AJ....106.2096V}.

During the W-D modeling process, the adjustable parameters are: the orbital inclination $i$; the mean surface temperature of star 2 ($T_{2}$); the monochromatic luminosity of star 1 ($L_{1}$); the dimensionless potential of star 1 ($\Omega_{1}$) and star 2 ($\Omega_{2}$), and the third light $l_3$. It is found the solutions converge at both Mode 3 (contact configuration) and Mode 5 (semi-detached configuration). The final photometric solutions are listed in Table 3. The theoretical light curves of Mode 3 is displayed in Fig. 2 and the theoretical light curves which haven't been contaminated by the third light are also plotted with dashed lines. The resulting figure from Mode 5 is very similar to Fig. 2. The geometrical structure at phase 0.25 is displayed in Fig. 3.

\begin{table}[!h]
\normalsize
\begin{center}
\caption{Photometric solutions of V921 Her.}
\begin{tabular}{lcc}
\hline
\hline
Parameter                     & solutions                      & solutions              \\\hline
Mode                          & 3                              & 5                       \\
$T_{1}(K)   $                 & 7700(assumed)                  & 7700(assumed)           \\
$g_{1}$                       & 1.00(assumed)                  & 1.00(assumed)         \\
$g_{2}$                       & 0.32(assumed)                  & 0.32(assumed)          \\
$A_{1}$                       & 1.00(assumed)                  & 1.00(assumed)          \\
$A_{2}$                       & 0.50(assumed)                  & 0.50(assumed)       \\
$q[m_{2}/m_{1}]$              & 0.226(assumed)                 & 0.226(assumed)   \\
$\Omega_{1}$                  & 2.296($\pm$0.015)              & 2.303($\pm$0.022)  \\
$\Omega_{2}$                  & 2.296($\pm$0.015)              & 2.296     \\
$\Omega_{in}$                 & 2.296                          & 2.296      \\
$i[^{\circ}]$                 & 75.4($\pm$3.8)                 & 75.5($\pm$3.5)          \\
$T_{2}[K]$                    & 7003($\pm$92)                  & 7011($\pm$84)      \\
$\Delta T(K)$                 & 697                            & 689                      \\
$T_{2}/T_{1}$                 & 0.91($\pm0.01$)                & 0.91($\pm0.01$)       \\
$L_{1}/(L_{1}+L_{2})$         & 0.869($\pm$0.013)              & 0.867($\pm$0.010)  \\
$L_{3}/(L_{1}+L_{2}+L_{3})$   & 0.188($\pm$0.074)              & 0.189($\pm$0.059)  \\
$r_{1}(pole)$                 & 0.478($\pm$0.004)              & 0.475($\pm$0.005)  \\
$r_{1}(side)$                 & 0.517($\pm$0.006)              & 0.514($\pm$0.007)  \\
$r_{1}(back)$                 & 0.541($\pm$0.007)              & 0.537($\pm$0.008)  \\
$r_{2}(pole)$                 & 0.241($\pm$0.005)              & 0.241    \\
$r_{2}(side)$                 & 0.251($\pm$0.005)              & 0.251      \\
$r_{1}(back)$                 & 0.283($\pm$0.009)              & 0.283  \\
$\Sigma res^{2}$              & $5.3659*10^{-6}$               & $5.3635*10^{-6}$  \\
\hline
\end{tabular}
\end{center}
\end{table}

\begin{figure}[!h]
\includegraphics[width=1.0\columnwidth]{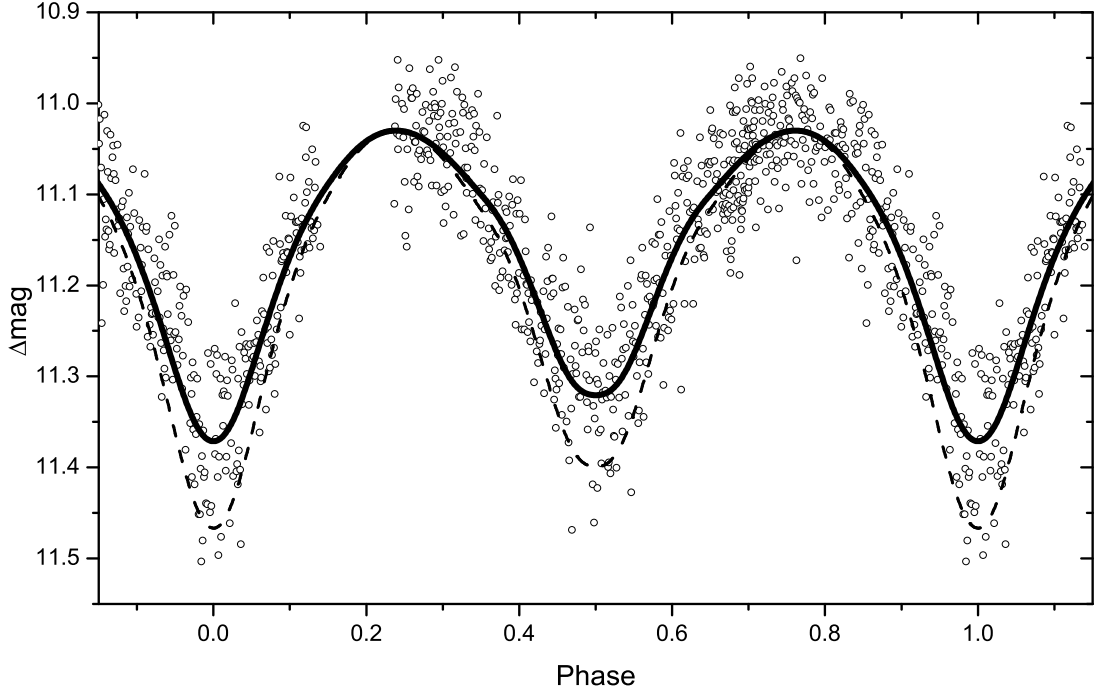}
\caption{\small The observed (open circles) and theoretical (solid line) light curve of V921 Her. Theoretical light curve without contaminated by the third light is plotted with dashed line. (Mode3)}
\end{figure}

\begin{figure}[!h]
\includegraphics[width=1.1\columnwidth]{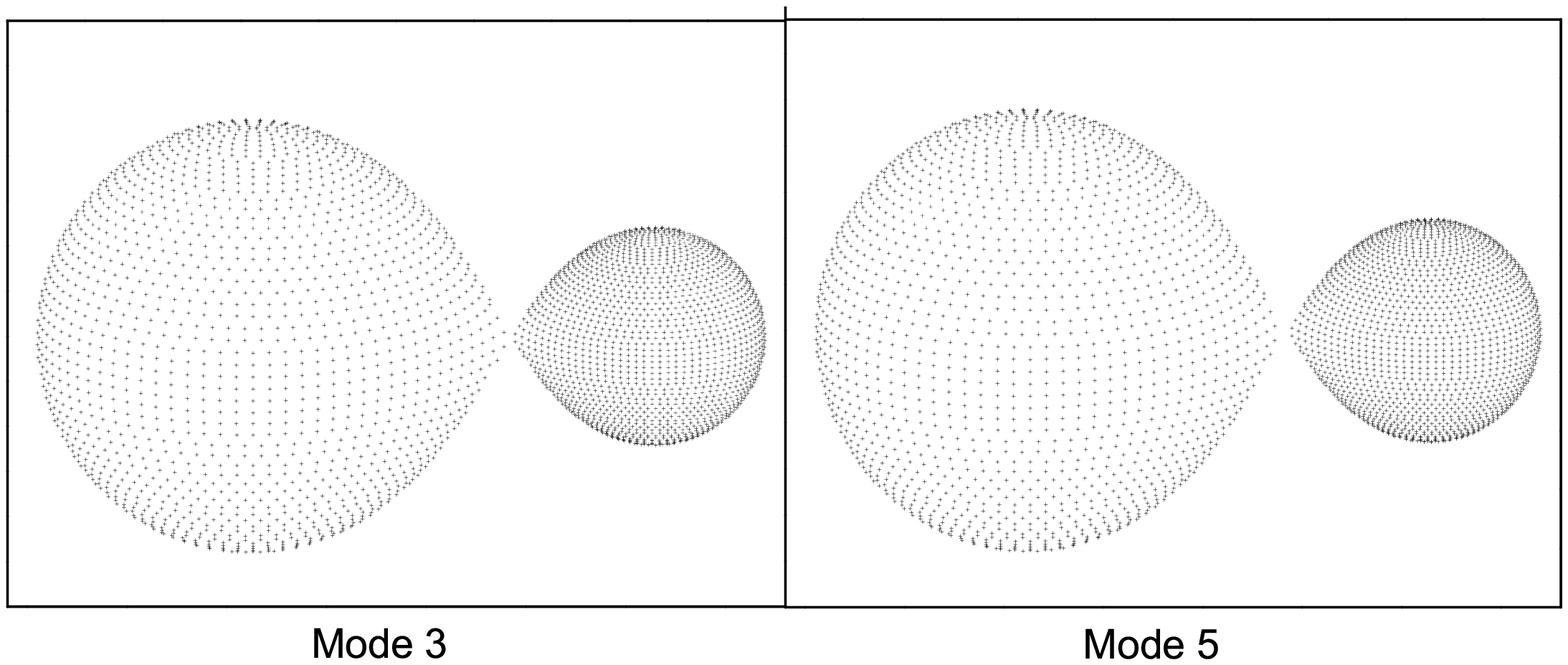}
\caption{\small Geometrical structure of V921 Her at phase 0.25.}
\end{figure}

\section{Discussion and Conclusions}
The light curve solutions of V921 Her show that it converges at both Mode 3 and Mode 5. The solutions of Mode 3 determines a contact configuration with both of its components just filling their critical Roch lobe, and the solutions of Mode 5 determines a semidetached binary system with the secondary component filling the critical Roch lobe and the primary one filling $98\,\%$ of the critical Roch lobe. However, as displayed in Table 3, the solutions between Mode 3 and Mode 5 have only very small differences in parameters such as orbital inclination $i$; temperature of star 2 ($T_{2}$); potential of star 1 ($\Omega_{1}$) and star 2 ($\Omega_{2}$); luminosity ratio of $L_{1}/L_{2}$; and the contribution of the third light $l_3$. Thus, we can conclude that V921 Her may be a marginal contact or B-type binary system \cite{2004A&A...426.1001C}. The temperature difference between its two components is about $700K$, which indicates that V921 Her is a poor thermal contact binary system. It is an important target which is exactly in the rare phase predicted by the thermal relaxation oscillation (TRO) theory \cite{1976ApJ...205..208L,1976ApJ...205..217F,1977MNRAS.179..359R,1979ApJ...231..502L}.

Compared with the solutions obtained by Gazeas et al. \cite{2006AcA....56..127G}, we determines a much higher orbital inclination and our temperature for star 2 is about $300K$ lower than theirs, which may due to the different mass ratio ($q=0.226$) we used. And also, we add a third light in our solutions, which may also account for these differences. As we conclude that V921 Her is a marginal contact binary system and being on the TRO stage, it is not so strange that a contact configuration was obtained by Gazeas et al. \cite{2006AcA....56..127G}.

Considering the orbital inclination ($i = 75^{\circ}$) of ours and the mass function given by Ruci{\'n}ski et al. \cite{2003AJ....125.3258R}: $(M_1+M_2)sin^3i = 1.971\pm0.061M_\odot$, we can easily calculate the mass of the two components to be $M_1 = 1.784(\pm0.055)M_\odot$, $M_2 = 0.403(\pm0.012)M_\odot$. And the orbital semi-major axis is obtained to be $a = 5.00(\pm0.15)R_\odot$. The absolute physical parameters of the two components in V921 Her are listed in Table 4. The absolute physical parameters are calculated according to the solutions of Mode 3. It is almost the same with the values calculated by using the solutions of Mode 5 since they give out nearly equal surface temperature, orbital inclination and so on.

\begin{table}[!h]
\caption{Absolute parameters of the two components in V921 Her}\label{absolute}
\begin{center}
\normalsize
\begin{tabular}{lllllllll}
\hline
Parameters                        &Primary                           & Secondary          \\
\hline
$M$                               & $1.784(\pm0.055)M_\odot$         & $0.403(\pm0.012)M_\odot$         \\
$R$                               & $2.56(\pm0.38)R_\odot$           & $1.29(\pm0.19)R_\odot$         \\
$L$                               & $12.2(\pm1.5)L_\odot$            & $1.59(\pm0.20)L_\odot$         \\
\hline
\end{tabular}
\end{center}
{\footnotesize Notes:} \footnotesize The absolute physical parameters are calculated according to the solutions of Mode 3.
\end{table}

As discussed by Liao \& Qian (2010)\cite{2010MNRAS.405.1930L}, the most plausible explanation of the cyclic period changes in close binaries is the light-travel time effect (LTTE) caused by the presence of the tertiary component around the binary system. By assuming a circular orbit ($e=0.0$), the projected radius of the orbit that the eclipsing binary rotates around the barycenter of the triple system is calculated with the following equation,
\begin{equation}
\begin{array}{lll}
a'_{12}\sin i'=A_3 \times c
\end{array}
\end{equation}
where $A_3$ is the amplitude of the $O-C$ oscillation and $c$ is the speed of light, i.e. , $a'_{12}\sin i'=0.89(\pm0.03)AU$. The mass function are computed with the following equation,
\begin{equation}
\begin{array}{lll}
f(m)=\frac{4\pi^2}{GP^2_3}\times(a'_{12}\sin i')^3=\frac{(M_3\sin i')^3}{(M_1+M_2+M_3)^2}
\end{array}
\end{equation}
where $G$ and $P_3$ are the gravitational constant and the period of the $(O-C)_2$ oscillation. Thus, we obtain the mass function to be $f(m)=0.007M_\odot$. The relationship between the orbital inclination ($i'$) and the mass ($M_3$) of the tertiary component are plotted in Fig. 4 (left panel). The relationship between the orbital inclination ($i'$) and orbital radius ($a_3$) of the tertiary component are also plotted in Fig. 4 (right panel).

\begin{figure}[!h]
\includegraphics[width=1.0\columnwidth]{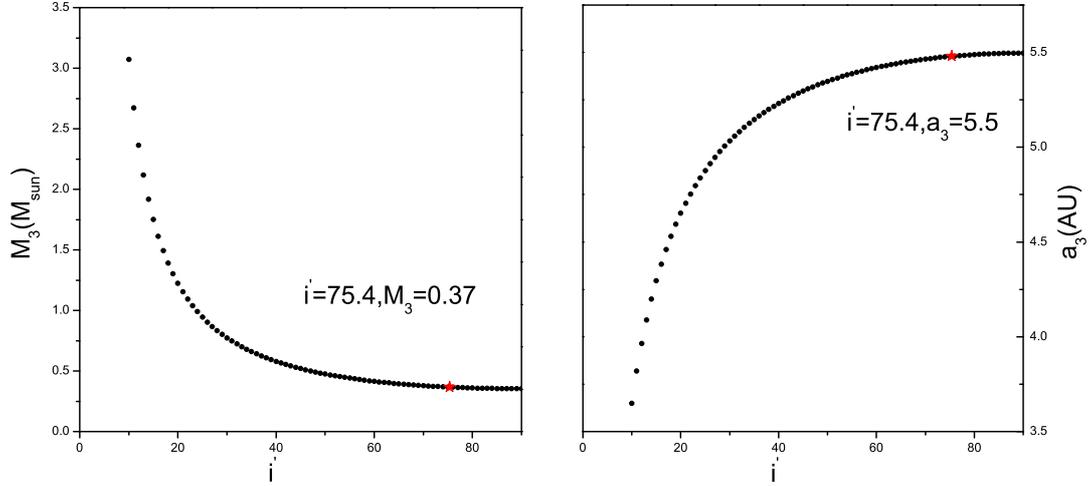}
\caption{\small The relationship between $i'$ and $M_3$ (left panel). The relationship between $i'$ and $a_3$ (right panel).}
\end{figure}

D'Angelo et al. \cite{2006AJ....132..650D} and Pribulla \& Rucinski \cite{2006AJ....131.2986P} pointed out that most overcontact binaries existed in multiple systems. A survey of 165 solar-type spectroscopic binaries concluded that the fraction of spectroscopic binaries with additional companions reached $96\,\%$ for $P < 3$d \cite{2006A&A...450..681T}. As for V921 Her, both light curve solutions and $O-C$ fitting results confirm the existence of a close-in tertiary component orbiting around the binary system. As show in Fig. 2, the third light significantly reduces the depth of the eclipse occultation. According to our results, the luminosity of the third component ($L_3$) is nearly twice of the luminosity of the secondary star ($L_2$). It is a stellar component which contributes about $19\,\%$ of the total luminosity in the triple system. It is supposed that the presence of a third component has played an important role in the formation and evolution of V921 Her by removing angular momentum from the central binary system during the early dynamical interaction \cite{2014AJ....148...79Q}.

The $O-C$ analysis also shows that the orbital period of V921 Her is increasing at a rate of $dP/dt=2.79\times{10^{-7}}day\cdot year^{-1}$. By considering a conservative mass transfer and combining the estimated mass of V921 Her with the well-known equation
\begin{equation}
\begin{array}{lll}
 \frac{dM_{2}}{dt}=\frac{M_1M_2}{3p(M_1-M_2)}\times{dp/dt}
 \end{array}
\end{equation}
the mass transfer from the less massive component to the more massive one, at a rate of $\frac{dM_{2}}{dt}=5.52\times{10^{-8}}M_\odot/year$ is determined.

V921 Her is under the key evolutionary stage predicted by the thermal relaxation oscillation (TRO) theory, which may evolve into a contact binary system. It is a very interesting target for testing theoretical scenario of formation and evolution on W UMa type binaries. More photometric and spectroscopic observations are still needed in the future. This object will be long time monitored.

\section*{Acknowledgements}
We thank the anonymous referees for their useful comments and suggestions that have improved the quality of our paper. We also appreciate the outstanding work of the LUT team and the support by the team from the ground system of the Chang¡¯e-3 mission. This work is supported by the Chinese Natural Science Foundation (Grant No. 11133007, 11325315 and 11203066), the Strategic Priority Research Program ``The Emergence of Cosmological Structure'' of Chinese Academy of Sciences (Grant No. XDB09010202), the Key Research Program of Chinese Academy of Sience (KGED-EW-603) and the Science Foundation of Yunnan Province (Grant No. 2012HC011).


\section*{References}
\begin{thebibliography}{9}
\bibitem{2014RAA....14.1511I} Ip, W.-H., Yan, J., Li, C.-L., \& Ouyang, Z.-Y.\ 2014, Research in Astronomy and Astrophysics, 14, 1511
\bibitem{1997ESASP1200.....E} ESA 1997, ESA Special Publication, 1200
\bibitem{2003AJ....125.3258R} Ruci{\'n}ski, S.~M., Capobianco, C.~C., Lu, W., et al.\ 2003, \aj, 125, 3258
\bibitem{2006AcA....56..127G} Gazeas, K.~D., Niarchos, P.~G., Zola, S., Kreiner, J.~M., \& Ruci{\'n}ski, S.~M.\ 2006, \actaa, 56, 127
\bibitem{2007JApA...28..217K} Karami, K., \& Mohebi, R.\ 2007, Journal of Astrophysics and Astronomy, 28, 217
\bibitem{2009AN....330..836K} Karami, K., Ghaderi, K., Mohebi, R., Sadeghi, R., \& Soltanzadeh, M.~M.\ 2009, Astronomische Nachrichten, 330, 836
\bibitem{2015RAA....15.1068W} Wang, J., Cao, L., Meng, X.-M., et al.\ 2015, Research in Astronomy and Astrophysics, 15, 1068
\bibitem{2015Ap&SS.358...47M} Meng, X.-M., Cao, L., Qiu, Y.-L., et al.\ 2015, \apss, 358, 47
\bibitem{1956BAN....12..327K} Kwee, K.~K., \& van Woerden, H.\ 1956, \bain, 12, 327
\bibitem{1997AAS...191.4815M} Marshall, S., Akerlof, C., Kehoe, R., et al.\ 1997, Bulletin of the American Astronomical Society, 29, 1290
\bibitem{2005IBVS.5668....1P} Pribulla, T., Baludansky, D., Chochol, D., et al.\ 2005, Information Bulletin on Variable Stars, 5668, 1
\bibitem{2007OEJV...74....1B} Br{\'a}t, L., Zejda, M., \& Svoboda, P.\ 2007, Open European Journal on Variable Stars, 74, 1
\bibitem{2005IBVS.5645....1P} Pejcha, O.\ 2005, Information Bulletin on Variable Stars, 5645, 1
\bibitem{2005IBVS.5602....1N} Nelson, R.~H.\ 2005, Information Bulletin on Variable Stars, 5602, 1
\bibitem{2005IBVS.5592....1K} Krajci, T.\ 2005, Information Bulletin on Variable Stars, 5592, 1
\bibitem{2005IBVS.5603....1D} Dvorak, S.~W.\ 2005, Information Bulletin on Variable Stars, 5603, 1
\bibitem{2007IBVS.5760....1N} Nelson, R.~H.\ 2007, Information Bulletin on Variable Stars, 5760, 1
\bibitem{2011IBVS.5974....1D} Dvorak, S.~W.\ 2011, Information Bulletin on Variable Stars, 5974, 1
\bibitem{2011IBVS.5966....1N} Nelson, R.~H.\ 2011, Information Bulletin on Variable Stars, 5966, 1
\bibitem{2011OEJV..137....1B} Br{\'a}t, L., Trnka, J., Smelcer, L., et al.\ 2011, Open European Journal on Variable Stars, 137, 1
\bibitem{1952ApJ...116..211I} Irwin, J.~B.\ 1952, \apj, 116, 211
\bibitem{Wilson1971} Wilson, R. E., Devinney, E.J. 1971, ApJ, 166, 605
\bibitem{Wilson1990} Wilson, R. E. 1990, ApJ, 356, 613
\bibitem{Wilson2010} Wilson, R. E., Van Hamme, W., Terrell, D., 2010, ApJ, 723, 1469
\bibitem{Cox2000} Cox, A. N. 2000, Allen$^\prime$s Astrophysical Quantities (4th ed.; NewYork: Springer)
\bibitem{1969AcA....19..245R} Ruci{\'n}ski, S.~M.\ 1969, \actaa, 19, 245
\bibitem{1967ZA.....65...89L} Lucy, L.~B.\ 1967, \zap, 65, 89
\bibitem{1993AJ....106.2096V} Van Hamme, W.\ 1993, \aj, 106, 2096
\bibitem{2004A&A...426.1001C} Csizmadia, S., \& Klagyivik, P.\ 2004, \aap, 426, 1001
\bibitem{1976ApJ...205..208L} Lucy, L.~B.\ 1976, \apj, 205, 208
\bibitem{1976ApJ...205..217F} Flannery, B.~P.\ 1976, \apj, 205, 217
\bibitem{1977MNRAS.179..359R} Robertson, J.~A., \& Eggleton, P.~P.\ 1977, \mnras, 179, 359
\bibitem{1979ApJ...231..502L} Lucy, L.~B., \& Wilson, R.~E.\ 1979, \apj, 231, 502
\bibitem{2010MNRAS.405.1930L} Liao, W.-P., \& Qian, S.-B.\ 2010, \mnras, 405, 1930
\bibitem{2006AJ....132..650D} D'Angelo, C., van Kerkwijk, M.~H., \& Ruci{\'n}ski, S.~M.\ 2006, \aj, 132, 650
\bibitem{2006AJ....131.2986P} Pribulla, T., \& Ruci{\'n}ski, S.~M.\ 2006, \aj, 131, 2986
\bibitem{2006A&A...450..681T} Tokovinin, A., Thomas, S., Sterzik, M., \& Udry, S.\ 2006, \aap, 450, 681
\bibitem{2014AJ....148...79Q} Qian, S.-B., Zhou, X., Zola, S., et al.\ 2014, \aj, 148, 79
\end{thebibliography}
\end{document}